\documentclass[preprint]{ptptex}
\usepackage[dvips]{graphicx}
\usepackage{amsmath,amssymb}

\makeatletter
\newcounter{figno}
\notypesetlogo
\preprintnumber[4cm]{EP-HOU09-002 \\ arXiv:0906.3938}
\title{ On Coherence Lengths of  Wave Packets } 
\author{ Kenzo~\textsc{ISHIKAWA} and  Yutaka~\textsc{TOBITA}}
\inst{
Department of Physics, Hokkaido University, \\
             Sapporo 060-0810, Japan }
\recdate{July 14,~2009;~Reviced September 9,~2009}

\abst{
The coherence lengths of one-particle states described using quantum wave
 functions are studied. 
We show that one particle states in various situations are not
 described using simple plane waves but using wave packets that 
are superpositions of plane waves. A wave packet is an approximate
 eigenstate of the free Hamiltonian and has a finite spatial size that
 we call the coherence length.  
 The coherence lengths in the coordinate space and momentum space
 are studied in this paper. We investigate  several mechanisms of
 forming  wave packets, stabilities of wave packets,  and
 transformations of wave packets. 
}

\begin{document}

\maketitle

%\c@figure=0
%\makeatother
%\pacs{ 73.43.Lp}
%\maketitle
%\newpage
\section{Introduction}

%\subsection{Why do we study wave packet ?}
When particles are identified using detectors, they show classical
trajectories. These classical trajectories are observed
because the particle's wave functions have finite sizes and the
probability for the particle to be 
observed within this width becomes  unity.

A position $x$ and a momentum $p$ satisfy the canonical commutation
relation,
\begin{eqnarray}
[x,p]=i \hbar,
\label{commutation-relation}
\end{eqnarray}
and 
a momentum eigenstate is extended in space. Thus, the momentum eigenstate
does not show a classical
trajectory, and a linear combination of momentum eigenstates, a wave
packet, which has a finite spatial size, shows a classical trajectory. 
Wave packets are necessary for
describing states with finite spatial extensions.

Wave packet behaves like a particle and is convenient for studying the
connection between quantum mechanics and classical
mechanics. Uncertainty relations and other properties of quantum
mechanics are described well using wave packets and
are explained in many textbooks of quantum mechanics \cite{yeazell}.
   
In scattering experiments, an overlap between the initial and
final states is studied. Since the final states are determined using
 detectors of finite sizes, they are described using wave  
packets. \cite{Goldberger}\tocite{Ishikawa-Shimomura} Their
sizes are normally semimicroscopic between microscopic and macroscopic 
lengths, and it is  good to 
approximate   the initial and
final states using plane waves if the
typical scales of targets and interaction lengths are microscopic.

The wave packets have been applied in various areas of physics such as
electromagnetic wave propagations\cite{Brillouin}, 
particle scatterings,\cite{Goldberger}\tocite{Sasakawa}
and neutrino oscillations.
\cite{Giunti}\tocite{Stodolsky} Moreover,
the fundamental problems of quantum mechanics that are connected with measurements
and their  implications in quantum information, entanglements, and others
are tested using beams of electrons, neutrons, and of photons or its
coherent states, lasers.  However, in these cases, either the wave is
classical or the coherence lengths
of waves in quantum physics are much larger than the typical scales of 
microscopic objects. The interactions of these waves with microscopic
objects were  studied using plane waves. A detailed qualitative 
study of microscopic physics, particularly of particle physics, has not
been made for wave packets.

We found that the situation has been changed now and there are many
occasions where the wave packet's effects are
important.\cite{Asahara}\tocite{Ishikawa-Tobita} Particularly in
the present problems of fundamental physics where high precision, high energy, long
distance, and other new circumstances are required, these effects are
expected to be important. Qualitative theoretical investigations of the
wave packet's effects are lacking. It is our objective to
study the deviations of one-particle states from simple plane waves and
their consequences in many-body quantum systems. We mainly study the
systems of relativistic invariance, where space and time are treated
equally.

The minimum wave packet is an idealistic wave packet that satisfies 
the minimum uncertainty relation
between the variances  of coordinates and momenta.
Although they have been studied often,\cite{Schiff,Schiff2} it is instructive
to review the properties of the minimum wave packet here for  later
convenience.
 
From the canonical commutation relation,
Eq.~$(\ref{commutation-relation})$,
uncertainties in the position and momentum satisfy
\begin{eqnarray}
\delta x \delta p \geq \frac{\hbar}{2}.
\label{uncertainty-product}
\end{eqnarray} 
In Gaussian wave packet, the coherent state of one  variable $x$,
\begin{eqnarray}
& &\langle x |P_0,X_0\rangle =N_1e^{i{ P_0 \over \hbar} (x-X_0)-{1 \over 2{\gamma}}(x-X_0)^2},\\
& &N_1^2=(\pi {\gamma})^{-{1 \over 2}},~{P_0 \over \hbar}=k_0, 
\end{eqnarray} 
where $X_0$ and $P_0$ are the expectation values of $x$ and $p$, the
equality of Eq.~$(\ref{uncertainty-product})$ is satisfied.  The 
variances of $x$ and $p$ are, in fact,
\begin{eqnarray}
& &( \delta x)^2=  \langle| x^2| \rangle-\langle |x |\rangle^2= {1 \over 2}{\gamma}^2,    \\
& & ( \delta p)^2= \langle| p^2| \rangle-\langle |p|\rangle^2  ={{\hbar}^2 \over {2 {\gamma}^2}}\label{parameter-sigma}
. 
\end{eqnarray}
The product of the variance of the momentum with that of the coordinate,   
\begin{eqnarray}
& &(\delta x)^2 \times (\delta p)^2= {(\hbar)^2 \over 4},  
\end{eqnarray} 
is independent from $\gamma $ and is  the minimum allowed from the 
commutation relation.   

The coherent state also satisfies the completeness condition,\cite{Ishikawa-Shimomura}
\begin{eqnarray}
\int {d P_0 d X_0 \over 2 \pi \hbar}\langle x| P_0,X_0\rangle \langle  P_0,X_0|y\rangle 
&=& \int {d P_0 d X_0 \over 2 \pi \hbar} N_1^2 e^{ik_0(x-y)}e^{-{1 \over 2 \gamma }(x-X_0)^2-{1 \over 2 \gamma }(y-X_0)^2 } \nonumber\label{completeness}\\
&=&\delta (x-y).
\end{eqnarray}
In higher dimensions, the products of the functions of each variable 
are used. They satisfy the minimum uncertainty relations and completeness 
conditions in higher dimensions.

For nonminimal wave packets, a function $ h_m(x-X_0) $ is multiplied to 
the right-hand side,
 \begin{eqnarray}
\langle x |P_0,X_0\rangle =N_1e^{i{ P_0 \over \hbar} (x-X_0)-{1 \over 2{\gamma}}(x-X_0)^2}h_m(x-X_0).
\label{non-minumum-wave packet}
\end{eqnarray} 
The completeness condition, Eq.~$(\ref{completeness})$, is satisfied also in 
this case, but the product of $\delta
x$ and  $\delta p$ is not minimum and is larger than the minimum. 
This packet  should be used  in  realistic cases of having a larger value
of the product of uncertainties. If the function $h_m(x-X_0)$ is
a Hermite  polynomial of order $m$, the product of uncertainties becomes
\begin{eqnarray}
\delta x \delta p=m \times {\hbar  \over 2 }\label{Harmonic}.
\end{eqnarray}

%\subsection{Observabilities of wave packets}

A wave packet emerges from the matter effects. The wave in the medium  is
affected by disorders and is described by one function within a finite
length, which we define as  a coherence length. When this wave of 
finite coherence is emitted from the medium  into the vacuum, it has 
a finite time length and a finite energy width, because 
the wave  in the matter lives for  a finite period of time, 
of the order of a mean free time.  
Consequently, this wave packet in the vacuum  is not the
eigenstate of the Hamiltonian but it is the linear combination of 
different energies.  Its width  is  determined from the 
interactions of the particle  with scatterers. There are other situations 
where wave packets are formed. We will study them and estimate wave
packet sizes in \S 2.

Since the wave packet is a linear combination of plane waves of different
energies, the wave packet is not a stationary state and varies with time. 
For a nonrelativistic particle of mass $m$,
the energy, $E={{\vec p}^2 \over 2m}$, has the  width  
\begin{eqnarray}
\Delta E={p \over m} \Delta p= v\Delta p,~v={p \over m},
\end{eqnarray}
and  the time width  is given by 
\begin{eqnarray}
\Delta \tau={\hbar \over \Delta E}={\Delta x \over v},~\Delta x={1 \over  \Delta k},~\Delta k={\Delta p \over \hbar }.
\end{eqnarray} 
In the above equations,  the energy width $\Delta E$, time width 
$\delta \tau$, and position width $\Delta x$ are determined using the 
momentum width $\Delta p$. 

The  position of the wave packet also varies  with time in a manner that
follows a classical trajectory of the velocity given by the central
value of the momentum. Furthermore, the wave packet spreads with time
with  a speed  that depends on the mass and initial size.

The other  feature of the wave packets can be observed in a momentum correlation,
\begin{eqnarray}
C({\vec p}_1,{\vec p}_2)=(\langle {\vec p}_1|\alpha \rangle )(\langle {\vec p}_2|\alpha  \rangle)^{*}=\langle {\vec p}_1|\alpha \rangle \langle \alpha|{\vec p}_2 \rangle,
\label{correlation-wave packet}
\end{eqnarray}
where the state $ | \alpha \rangle $ is one state described using a wave
packet and $ | {\vec p} \rangle $ is a momentum eigenstate. If the 
state $| \alpha \rangle $ is 
a momentum state $|{\vec q} \rangle $, then we have 
\begin{eqnarray}
C_0({\vec p}_1,{\vec p}_2)=(\langle {\vec p}_1|\vec q \rangle )(\langle {\vec p}_2|
{\vec q}'  \rangle)^{*}=\delta ({\vec p}_1-{\vec q})\delta ({\vec p}_2-{\vec q}),
\end{eqnarray}
and  $C_0({\vec p}_1,{\vec p}_2)$ is proportional to $\delta({\vec
p}_1-{\vec p}_2)$. 

 If the state $| \alpha \rangle $ is 
a wave packet described using the momentum state $|{\vec q} \rangle $ as
\begin{eqnarray}
| \alpha \rangle=\int d{\vec q} F_{\alpha}({\vec q})|\vec q \rangle,
\end{eqnarray}
then we have 
\begin{eqnarray}
C({\vec p}_1,{\vec p}_2)&=&\int d{\vec q}_1 F_{\alpha}^{*}({\vec q}_1) 
d{\vec q}_2 F_{\alpha}({\vec q}_2)\delta({\vec p}_1-{\vec q}_1)\delta({\vec p}_2-{\vec q}_2)\\
&=& F_{\alpha}^{*}({\vec p}_1) F_{\alpha}({\vec p}_2).
\nonumber
\end{eqnarray}
Thus, the correlation function vanishes if $|{\vec p}_1-{\vec p}_2|$ is larger 
than the momentum width of the wave packet. When   all 
the states of a complete set are added,
\begin{eqnarray}
\sum_{\alpha} C({\vec p}_1,{\vec p}_2)=\sum_{\alpha } \langle {\vec p}_1|\alpha \rangle \langle \alpha|{\vec p}_2 \rangle
=\delta({\vec p}_1-{\vec p}_2),
\end{eqnarray}
then the correlation function agrees with the delta function. For a state 
described using a wave packet, $C({\vec p}_1,{\vec p}_2)$ deviates 
from $\delta({\vec p}_1-{\vec p}_2)$.
The
deviation of the correlation function $C({\vec p}_1-{\vec p}_2)$ from the 
$\delta({\vec p}_1-{\vec p}_2)$ shows a feature of the state $|{\alpha }
\rangle$, and the width of ${\vec p}_1-{\vec p}_2$ is determined using the
momentum width $\Delta p$. This correlation function will be used later.

 We investigate the problems  connected with the particle's coherence  in 
this series of papers. In  a previous paper,\cite{Ishikawa-Shimomura}
we showed the general features of wave packet scatterings such as the 
evolution of wave packets and slight violations of energy and momentum
conservations in many-body reactions. The consistency of the 
nonorthogonality of wave packets with the fundamental
requirement of quantum mechanics in many-body scatterings was also shown.  

In  the present paper, we study the formations of  wave packets 
that have finite uncertainties of position and momentum and the 
transmutations of these uncertainties in various reactions. 
We present the coherence length of wave packets  and other universal
properties  in  the potential scatterings and other many-body reactions.

This paper is organized as follows. In \S2, several
mechanisms of forming wave packets are studied. In \S3, we
study the potential scatterings of wave packets, and in \S4,
we present the transformations of wave packets.  In \S5 the particle's
coherence in refraction and reflection is studied, and  many-body processes 
are studied in \S6. A summary is given in \S7.
%\newpage

\section{Wave packet formations}
In this section, we study one-particle states in various situations and
show that the wave packets are good wave functions for particles in
medium and for particles in measurements. 

First, we study a system where the particle's mean free path is short. If
a particle interacts with atoms or other particles frequently, the
distance in
which a particle moves freely, the mean free path, 
becomes short. 
A particle is described using one wave function
during  a mean free time, which is a period for a particle to move
freely, and is given by dividing the mean free path 
by its velocity.   This wave function that has a finite coherence length
is described using a wave packet. 

Second, we study one  particle surrounded by many particles where the particle's
mean free path is long. 
In this system, particles have long mean free paths, and the direct
effects of the
mean free path are negligible. However, the  many-particle state is described using one 
wave function.  If a one-particle state in a many-particle system is regarded
as a linear combination 
of momentum states, this particle is expressed using a wave packet. 
This wave packet has  a different origin from that of the first
case  and plays important roles  in dilute medium.

Third, we study a one-particle state in a system where a particle measurement
 is made. If a
particle is measured with uncertainties of position and momentum, this
particle state is described using a wave packet of these uncertainties. 
The formation of wave packets in the process of particle measurement is a delicate problem that 
is connected with the fundamental problem of measurement of quantum 
mechanics. In fact, a particle is identified using a classical trajectory in
a  detector, and its position and
momentum are measured with finite uncertainties regardless of the
dynamics  of measurement. Thus, the state that has these uncertainties of
momentum and position is described using a wave packet.    

\subsection{Short mean free path:~finite spatial extensions}

Particles in matter frequently interact with atoms and lose coherence. 
The average distance for one particle to move freely is the mean free path 
in which the particle's  wave maintains coherence. Beyond the mean free path,
particles lose coherence and are expressed using different wave
functions. Hence, this state has a finite spatial width and its momentum
is defined with a finite uncertainty. This momentum uncertainty is inversely
proportional to the mean free path and becomes large if the mean free
path is short.  Thus, a mean free path is used as a  
wave packet size. 

\subsubsection{Mean free path }
The mean free path of a particle when it propagates 
in matter and is scattered incoherently by scatterers is determined 
using the cross section and  number density of scatterers. 
From the scattering cross section of a particle, $\sigma$, 
and number density of the scatterers, $\rho$, the mean free path $l$
is determined as
\begin{eqnarray}
l={1 \over \sigma \rho }.
\end{eqnarray} 
The mean free paths of various particles in matter are computed easily.

\subsubsection{Comparison of energy widths  }

Because the particle state is defined using one wave function within the mean free
path, $l$, this state   has a finite uncertainty of the 
momentum, $\delta p$,
\begin{eqnarray}
\delta p=\frac{\hbar}{l}.
\end{eqnarray} 
This finite uncertainty  of the momentum leads to a finite uncertainty of
the energy for the nonrelativistic particle of mass $m$,
\begin{eqnarray}
\delta E={\delta p^2 \over 2m}=v\delta p=v\frac{\hbar}{l}.
\label{delta-energy-1}
\end{eqnarray} 
Thus, the particle that has a mean free path $l$ has the energy width 
$v{\hbar \over l}$. 

The uncertainty of the energy of a wave packet is found from the momentum width  
Eq.~$(\ref{parameter-sigma})$,
\begin{eqnarray}
\delta E={\delta p^2 \over 2m}=v\delta p=v{\hbar \over \sqrt{ 2}\gamma}.
\label{delta-energy-2}
\end{eqnarray} 
Comparing two energy widths, Eqs.~$(\ref{delta-energy-1} )$ and $(\ref{delta-energy-2} )$, we have 
the wave
packet parameter $\gamma$,
\begin{eqnarray}
\gamma= {l \over \sqrt 2}.
\end{eqnarray} 
Thus, the wave packet size $\gamma$ is determined using the mean free path $l$.

When this particle moves with a velocity $v$,
the time spent by this particle crossing one position is
given using the mean free path over the particle's velocity,
\begin{eqnarray}
\tau ={l \over v  }.
\end{eqnarray}  
This state has an uncertainty of time $\tau$ and an uncertainty of 
energy $\delta E$.  $\delta E$ is given by
\begin{eqnarray}
\delta E={\hbar \over \tau}={\hbar \over l}v.
\label{delta-energy-3}
\end{eqnarray}  
The energy width of Eq.~$(\ref{delta-energy-3})$ agrees with those of   
Eqs.~$(\ref{delta-energy-1})$ and $(\ref{delta-energy-2})$.

Thus, in matter, a particle wave has a finite spatial size. Consequently, a wave 
behaves as a wave packet of this spatial size.
The momentum width is determined from the inverse of the spatial width, and
the wave is approximately given using the minimum wave packet.

\subsection{Long mean free path:~finite momentum spreads}
When there is a  degeneracy in one particle energy, a superposition
of the states with the same energy is also an eigenstate.
Which  eigenstate is realized depends on the production process.  
In this section, we study wave packets that have origins in momentum
spreading of the produced particles and are connected with the energy 
degeneracy.  

\subsubsection{Many-particle states:~transmutation of momentum spreads}
In a system of many particles with long mean free path, one particle  
interacts with other surrounding particles  many times while maintaining  
quantum coherence.  If these surrounding particles have momentum
uncertainties, they are transmuted to one particle.    This particle is
described using a wave packet that has an origin  in the momentum 
uncertainty.    

One example of a wave that is extended in the momentum is a spherical
wave around a short-range potential.  A spherical wave is a
superposition of plane waves and is decomposed into 
plane waves of continuous momentum. Hence, the spherical wave is
extended in  the momentum. In a normal scattering, a particle is
measured at a certain scattering angle. The probability of observing this 
particle at a certain angle is proportional to the square of the absolute
value of the amplitude.  It is not easy  to verify directly the fact
that the wave function   is extended in  momentum variables. In our
objective of studying interference phenomena in systems of large scales,
this type of momentum extension is  important as  a new mechanism of  
the wave packet.

We study the coherence length of a particle surrounded by many particles 
and whole states are described using one wave function based on 
the correlation in the momentum variables $C({\vec p}_1,{\vec p}_2)$. In 
this  situation, one particle obtains a large
uncertainty from many particles. Even though the momentum uncertainty of
each particle is   small, the effects of momentum uncertainties of these
many particles are added constructively and affect this particle with a 
significant magnitude. Consequently, this particle is given a finite 
momentum uncertainty by these many particles and behaves as a wave packet.

This new mechanism is applied, for instance,  when many particles are
involved in the microscopic processes and they maintain coherence for a long time.
This may be realized actually in the universe.  Particularly
around the decoupling time of cosmological background 
radiations in an early universe,  
where particle states are described using one wave function for a long
time, a wave packet due to momentum uncertainty  of this section is 
expected to play roles. 

During the time evolution of one particle wave function,  this particle
interacts with surrounding  particles. The effects of surrounding
particles are studied next. We study the system of many electrons and 
many photons in which they interact by  Thomson scattering. The Thomson 
scattering  cross section is given in Appendix A. 
%\begin{eqnarray}
%\sigma_{Thomson}={ 8 \pi r_e^2 \over 3},r_e=
%{e^2 \over 4\pi \epsilon_0 m_e c^2}
%\end{eqnarray}
Photons  and electrons  interact also with protons. The number of protons is 
the same as
the number of electrons from charge neutrality. These protons
have finite uncertainties, i.e., finite spreads  of the momenta due to 
Rutherford scatterings between charges
\begin{eqnarray}
\Delta p={\hbar \over l_{\text{Ru}}},~l_\text{Ru}={1 \over \sigma_{\text{Ru}} n_\text{c}},
\end{eqnarray}
where $n_c$ is the density of charged particles and $\sigma_\text{Ru}$ is
Rutherford scattering cross section, which is given in Appendix A. 

Even 
if  these momentum
spreads   are small, a system of many 
protons could give a large uncertainty to an electron owing to the  
constructive effects.   In this case, the whole uncertainty of
the momentum  becomes large, and this system of the electron and 
protons violates the translational invariance maximally.

A mean free path of the electron due to Thomson scattering is given by
\begin{eqnarray}
l_\text{Th}={1 \over \sigma_\text{Th} n_{\gamma}},
\end{eqnarray}
where $n_{\gamma}$ is photon density.  A mean free path of the
electron due to Rutherford scattering is given by
\begin{eqnarray}
l_\text{Ru}={1 \over \sigma_\text{Ru} n_\text{pr}},
\end{eqnarray}
where $n_\text{pr}$ is a proton density.  Their magnitudes are given in
Appendix A.
The ratio between two lengths
\begin{eqnarray}
N_\text{T}={l_\text{Th} \over l_\text{Ru} }
\end{eqnarray}
is the number of average collisions due to Rutherford scattering per
Thomson scattering. During a Thomson scattering time, Rutherford
scatterings occur $N_\text{T}$ times.  The $l_\text{Th}$, $l_\text{Ru}$, and $N_\text{T}$ are
given in Appendix A.  

\begin{figure}[t]
 \centerline{
 \includegraphics[scale=.5]{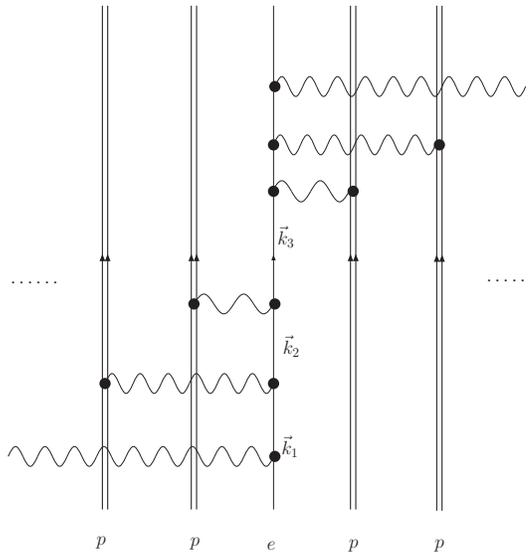}
 }
\refstepcounter{figno}\label{fig:proton-electron}
\caption{Feynman diagram in which an electron interacts with
 surrounding protons.}
\end{figure}

For the coherence of photon and electron, 
surrounding protons  are taken into account, and the Feynman diagram of 
these processes has many lines of 
particles, as in Fig.~\ref{fig:proton-electron}. 
%\vspace{7cm}
%%%%%%%%%%%%%%%%%%%%%%%%%%%% Fig 1 %%%%%%%%%%%%%%%%%%%%%%
%%%%%%%%%%%%%%%%%%%%%%%%%%%% Fig 1 %%%%%%%%%%%%%%%%%%%%%%
The average number of Rutherford scatterings in the
distance $l$, $N$,  is given by 
\begin{eqnarray}
N={l \over l_\text{Th}}.
\end{eqnarray}
Thus, the electron's spread obtains contributions from the proton's momentum
spread  $N$ times in the distance $l$. A photon obtains contributions from
the electron's momentum spread.
As shown in Appendix B, the
momentum spreads  of the electron and photon, after the $N$ step of interactions, are given by
\begin{eqnarray}
(\Delta p_\text{total})^2=\sum_i^N (\Delta p_i)^2,
\end{eqnarray}
and its magnitude becomes large  for large  $N$. Particularly if 
$(\Delta p_i)^2=(\Delta p)^2$ is finite, $(\Delta p_\text{total})^2$ is given 
by 
\begin{eqnarray}
(\Delta p_\text{total})^2=N (\Delta p)^2
\end{eqnarray}
and reaches the absolute value of the momentum,~$|p|$,  at a sufficiently 
large  $N$. 
This  is realized 
at a macroscopic distance $l$.   Thus, the total  
spread  becomes 
\begin{eqnarray}
\Delta p_\text{total} \rightarrow \infty,
\end{eqnarray} 
and the wave has a  large momentum spread and the momentum conservation 
becomes effectively negligible.

\begin{figure}[t]
 \centerline{
 \includegraphics[scale=.5]{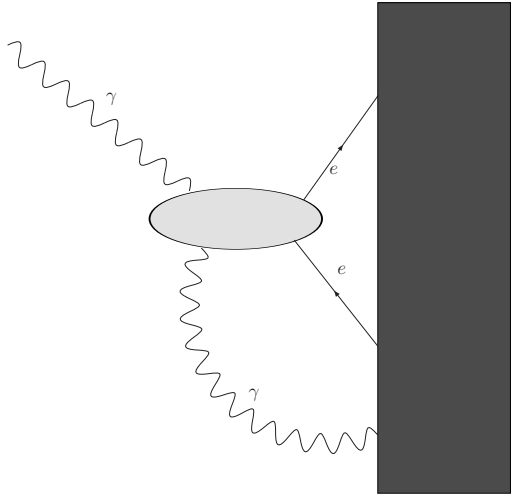}
 }
\refstepcounter{figno}\label{fig:stat}
\caption{Final Thomson scattering in which one photon and one electron
 interact and the photon in the final state is detected. The black box shows
 many-particle states. }
\end{figure}

Next, we focus the final Thomson scattering of one photon and one
electron shown in Fig.~\ref{fig:stat}. The incoming  states of the final scattering are almost real
photon and electron and have large momentum spreads. 
These  photon and electron  are described by  
superpositions of momentum states of large momentum uncertainties with a 
suitable energy weight. 

%%%%%%%%%%%%%%%%%%%%%%%%%%%% Fig 2 %%%%%%%%%%%%%%%%%%%%%%
%%%%%%%%%%%%%%%%%%%%%%%%%%%% Fig 2 %%%%%%%%%%%%%%%%%%%%%%

%\vspace{7cm}
\subsubsection{Statistical model}
The weight of the superposition of the initial states at the final
scattering is determined using its
previous scatterings and identical particle effects.
From the facts that the waves have large momentum spreads after the 
macroscopic distance and that the amplitude of the 
Thomson scattering in low energy of the range 3000 - 4000 K is constant and spherically symmetric,
 waves of the 
photon and electron in the initial state of the final scattering are 
regarded as spherical waves that are superpositions of plane waves
of all orientations.

In the many-particle state, identical  particles satisfy either Fermi-Dirac
statistics or Bose-Einstein statistics, and the total energy conservation law 
\begin{eqnarray}
\sum_i^N n_i E_i=E,
\end{eqnarray}
is satisfied. An average occupation number of the state ${\vec p}$, i.e., one 
particle distribution, is given by Bose-Einstein distribution 
\begin{eqnarray}
n_\text{P}({\vec p})=N_\text{BE}({\vec p})={1 \over e^\frac{p^0}{k_\text{B} T}-1}
\end{eqnarray}
for photon   and by Fermi-Dirac distribution 
\begin{eqnarray}
n_e({\vec p})=N_\text{FD}({\vec p})={1 \over e^{{p^0-\mu} \over k_\text{B} T}+1}
\end{eqnarray}
for electron, where $\mu$ is the chemical potential, $k_\text{B}$ is Boltzmann
constant, and $T$ is the temperature that is determined from the average 
energy. A many-body wave function that satisfies these conditions mentioned
above is the coherent state of the Boson field and the state of the Fermi field
constructed in the same manner. An explicit form is given in Appendix C. 

Thus, we assume
that the amplitude for 
one   photon and one electron is given by  
\begin{eqnarray}
T(1)= N_\text{BE}(p_2)^{1/2}N_\text{FD}(k_2)^{1/2}\tilde T(1),
\end{eqnarray}
where 
%$N_F(p_2)$ is the Fermi-Dirac distribution function and $N_B(k_2)$ is
%the Bose-Einstein distribution function(Fermi-distribution), and 
$\tilde T(1)$ does not depend on ${\vec p_2}$
and ${\vec k_2}$. The  whole amplitude is written as  
 \begin{eqnarray}
T({\vec k}_1,{\vec p}_1)&=&  \int d{\vec k}_2 d{\vec p}_2 T({\vec p}_1,{\vec k}_1;{\vec p}_2,{\vec k}_2)N_\text{F}(p_2)^{1/2}N_\text{B}(k_2)^{1/2}
\delta({\vec p}_2+{\vec k}_2-{\vec p}_1-{\vec k}_1)\nonumber \\
& &\times \tilde T(1),
\end{eqnarray}
where $T({\vec p}_1,{\vec k}_1;{\vec p}_2,{\vec k}_2)$ is the 
amplitude of Thomson scattering and is  constant in low
energy.

The momentum correlation function $C({\vec p}_1,{\vec p}_2)$ is a
product  of the amplitude of the photon momentum ${\vec p}_1$ and 
its complex conjugate of the photon momentum ${\vec p}_2$ 
and is  given by
\begin{eqnarray}
& &C({\vec k}_1,{\vec k'}_1)= 
 \int d{\vec k}_2 d{\vec p}_2 T({\vec p}_1,{\vec k}_1;{\vec p}_2,{\vec k}_2)N_\text{F}(p_2)^{1/2}N_\text{B}(k_2)^{1/2}\delta({\vec p}_2+{\vec k}_2-{\vec p}_1-{\vec k}_1)\nonumber\\
& &\int d{\vec k}'_2 d{\vec p}'_2  T({\vec p}'_1,{\vec k}_1;{\vec p}'_2,{\vec k}'_2)N_\text{F}(p'_2)^{1/2}N_\text{B}(k'_2)^{1/2}\delta({\vec p}'_2+{\vec k}'_2-{\vec p}_1-{\vec k}'_1)|\tilde T|^2.
\end{eqnarray}
This correlation $C({\vec k}_1,{\vec k'}_1)$  shows the wave packet
nature of photons.
 
\begin{figure}[t]
 \includegraphics[scale=.4,angle=-90]{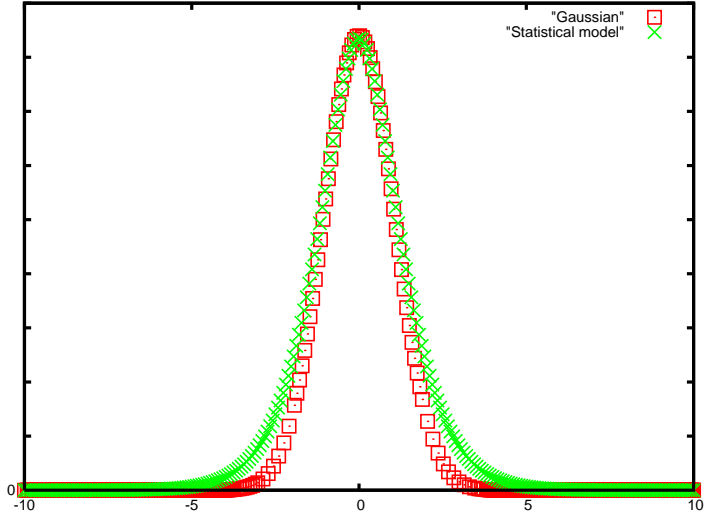}
\refstepcounter{figno}\label{3500K}
\caption{The photon correlation function of a statistical model shown by
 crosses is
 compared with a Gaussian function shown by boxes. The width of the Gaussian
 function is about 3.5$k_{B}T$ and  $T=3500$K.}
\end{figure}

We compute the above function $C({\vec k}_1,{\vec k'}_1)$ numerically.  
The result is given in Fig.~\ref{3500K}.    
%%%%%%%%%%%%%%%%%%%%%%%%%%%% Fig 3 %%%%%%%%%%%%%%%%%%%%%%
%%%%%%%%%%%%%%%%%%%%%%%%%%%% Fig 3 %%%%%%%%%%%%%%%%%%%%%%

%\vspace{5cm}
As seen from Fig.~\ref{3500K}, the correlation shows that of a Gaussian wave
packet of the width of $3.5k_\text{B}T$. The photon is regarded as a wave packet
whose energy distribution is a Planck distribution but the momentum width
is $3.5k_\text{B}T$. This wave packet size should be understood as 
a maximal possible value. The effect of the wave packet is show in Ref.~\citen{Ishikawa-Tobita}.

\subsection{Measurements of particle's trajectory}

When a particle is measured using an apparatus, its position and momentum
are measured within
certain uncertainties.    Many physical processes are
involved in measurements, but regardless of these  processes when  the
position and momentum are determined within uncertainties, 
the final state after the measurement is expressed as the state with
these  uncertainties. Thus, a wave packet of  suitable values of
momentum width and coordinate width is used for this state. The
probability for the particle to be observed within these widths of wave
packet is unity, and it follows a classical motion before the next measurement.

The product of uncertainties of the momentum $\Delta {\vec p}$ and
position $\Delta {\vec x}$ may be much
larger than that of the minimum wave packet, 
\begin{eqnarray}
|\Delta {\vec x}|| \Delta {\vec p}| \gg {\hbar \over 2},
\end{eqnarray}
so nonminimum wave packets $\left(\text{Eq}.~(\ref{non-minumum-wave packet})\right)$
are suitable for these states.

By successive measurements, a particle's trajectory that follows
classical motion is observed \cite{Schiff2}. This happens because the wave
packet has a finite spatial size, and the probability for this particle to
be observed in the inside of this region is unity.     
Actually, as discussed in a previous paper,\cite{Ishikawa-Shimomura}
the wave packet spreads with
time. 
The spreading velocity depends on the mass and energy. For the
particle trajectory to be observed, a next successive reaction with
an apparatus should occur before the wave becomes large. Unless
an observation is made, the wave spreads ultimately and a straight
trajectory is not tested.

Because light microscopic particles spread fast, they become momentum
eigenstates easily. Hence, translational invariance is preserved for these
particles. On the other hand, for macroscopic objects or extremely heavy
particles, the spreading velocities are negligible and they are localized at
certain positions of the initial states. Translational invariance is
violated for these objects.

\section{Potential scatterings }
In this section, we study the scatterings of the wave 
packets using simple rectangular potentials. We see that the sizes of extensions 
in the positions and momenta have strong correlations with the
velocities of the wave packets and the products of both sizes are 
approximately adiabatic invariants. Minimum wave packets are changed to 
nonminimum  wave packets in certain reactions.

\subsection{Potential wall in one dimension}
%\subsubsection{Steep  wall}

We study the wave packets first in a simple potential, i.e., in a constant 
potential wall of a height or depth $V_0$ 
described by 
\begin{eqnarray}
U(x) = \begin{cases}
	V_0, \ x \leq 0, \\
	0, \ \ 0 < x.
       \end{cases}
\end{eqnarray}
We obtain a solution  of Schr\"{o}dinger equation,
\begin{eqnarray}
& & i\hbar {\partial \over \partial t} \psi(x,t)=H\psi(x,t),\\
& &H={p^2 \over 2m}+U(x),
\end{eqnarray}
of the following form,
\begin{eqnarray}
\psi(x,t)&=&e^{Et \over i\hbar} \psi(x), \\
\psi(x) &=& \begin{cases}
	   e^{i k x}+ B_{-}e^{-i k x},~x < 0, \\     
	   C_{+}e^{i k'x},~0 < x,
	  \end{cases}
\end{eqnarray}
 where at $x<0$, a right-moving plane wave comes in toward the wall and
a 
left-moving wave
of magnitude $B_{-}$  is reflected, and at $x>0$, a right-moving wave of 
magnitude $C_{+}$ is refracted.   The parameters $k$ and $k'$ are
connected with the energy as
\begin{eqnarray}
& &E={p^2 \over 2m}+V_0={p'^2 \over 2m},\\
& &p=\hbar k,~p'=\hbar k'.
\end{eqnarray}
The coefficients are found from the continuities of the wave function at
$x=0$ as \cite{Landau}
\begin{eqnarray}
& &B_{-}={k-k' \over k+k'},\\
& &C_{+}={2k \over k+k' }.
\end{eqnarray}
The time-dependent wave packet is constructed using a linear combination of the above waves,
\begin{eqnarray}
\psi_{wp}&=&\psi_{wp}^1+\psi_{wp}^2+\psi_{wp}^3,\\
\psi_{wp}^1&=&\int dk \tilde N(k) e^{i k (x-X_0)}e^{Et \over i\hbar}, x<0,       \\
\psi_{wp}^2& &=\int dk \tilde N(k)\left( {k-k' \over k+k'}\right) e^{-i k(x-X_0)}e^{Et \over i\hbar},x<0  ,\\
\psi_{wp}^3& &=\int dk \tilde N(k)\left({2k \over k+k'}\right)e^{i k'(x-X_0)}e^{Et \over i\hbar},0<x ,
\end{eqnarray}
where 
\begin{eqnarray}
\tilde N(k)&=&N_1 e^{-{(k-k_0)^2 \over 2\sigma}},\\
\label{normalization}
N_1&=&{1 \over \sqrt{2\sigma \pi}}.
\end{eqnarray}
%%%%%%%%%%%%%%%%%% modefied %%%%%%%%%%%%%%%%%%%%%%%%%%%%%%%%%%%%%%%%%%%%%%%%%%%%%%%%%%%%%%%%%%%%%%%%%

\begin{figure}[t]
\includegraphics[scale=.4,angle=-90]{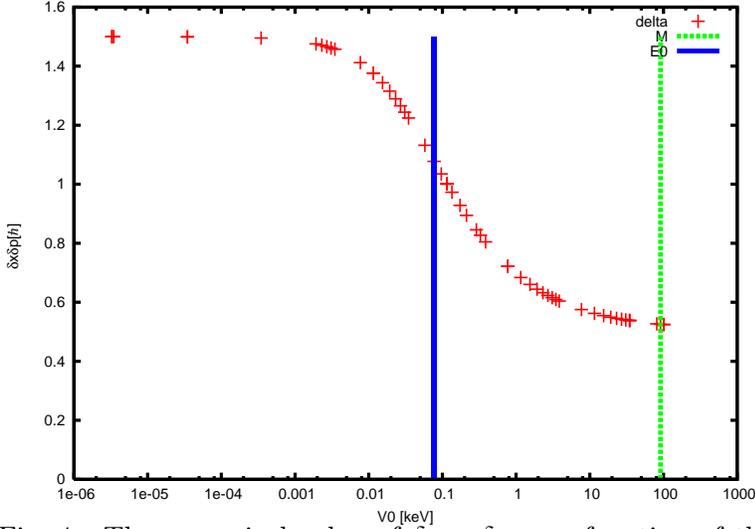}
\refstepcounter{figno}\label{cliff}
\caption{The numerical value of $\delta
 x\times \delta p$ as a function of the potential depth, $V_0$, for
 accelerated 
wave packet is given. The horizontal axis shows  the potential depth and
 the vertical axis shows $\delta x \times \delta p$. ``delta'' is for a numerical
 result, $E_0$ is average energy and M is mass.}
\end{figure}

The wave packet $\psi_{wp}^1$ is a minimum wave packet, but the wave
packets  $\psi_{wp}^2$ and $\psi_{wp}^3$ are not minimum wave packets
owing to the momentum-dependent factors in the amplitudes, ${{ k-k' \over
k+k'},{2k \over k+k'}}$. 
%The variances
%of  $\psi_{wp}^1$, $\psi_{wp}^2$, and $\psi_{wp}^3$ are computed
%numerically and are given by 
\begin{eqnarray}
& &\delta x^1 \times \delta p^1= {\hbar \over 2},\\
& &\delta x^2 \times \delta p^2 = (1.2 \sim 1.4)\times {\hbar \over 2},\\
& &\delta x^3 \times \delta p^3= (1.0 \sim 3.0)\times {\hbar \over 2}.
\end{eqnarray}
The initial wave packet is chosen to be minimum and satisfies the minimum
uncertainty relation, and the reflected wave and accelerated wave 
have  uncertainty relations of about twice the minimum. 
In particular, the value of $\delta x \delta p$ for the accelerated wave
depends on the potential depth, as given in Fig.~{\ref{cliff}}. The product of
uncertainties becomes $\frac{1}{2}\hbar$ at $V_0 \rightarrow \infty$ and $\frac{3}{2}\hbar$ at
$V_0 \rightarrow 0$. This is because the wave function behaves in
these regions as  
\begin{eqnarray}
 \psi_{wp}^3 \propto
  \begin{cases}
   k\times e^{-\frac{k^2}{2\sigma}} ,\ \ V_0 \ll E_0,\\
% ,~\text{then}~\delta x \delta p \to \frac{3}{2}
   \text{const}\times e^{-\frac{k^2}{2\sigma}}, \ \   V_0 \gg E_0,
%,~\text{then}~\delta x \delta p \to \frac{1}{2}
  \end{cases}
\end{eqnarray}
just like a wave function for a ground state and a first excited
state of a harmonic oscillator in Eq.~(\ref{Harmonic}).
The product of uncertainties of the accelerated wave is changed smoothly  as
seen in Fig.~\ref{cliff}.  
%Thus $\delta p \times \delta x$ is approximately adiabatic
%invariant in the present process of the wave packet.
 
%%%%%%%%%%%%%%%%%%%%%%%%%%%% Fig 4 %%%%%%%%%%%%%%%%%%%%%%
%%%%%%%%%%%%%%%%%%%%%%%%%%%% Fig 4 %%%%%%%%%%%%%%%%%%%%%%

\subsection{Potential barrier }
%\subsubsection{Rectangular barrier}
We study wave packets next  in a simple potential, i.e., in a constant 
potential barrier of finite width $a$ and height or depth $V_0$ 
described as
\begin{eqnarray}
U(x)=\begin{cases}
      0, ~x \leq 0, \\
      -V_0 ,~     0 \leq x \leq a,\\
      0, ~a\leq x.
     \end{cases}
\end{eqnarray}
We obtain a wave function of the following form:
\begin{eqnarray}
\psi(x)=
\begin{cases}
 e^{i{1 \over \hbar}\sqrt{2mE}x}+ B_{-}e^{-i{1 \over \hbar}\sqrt{2mE}x},~x<0,\\
=A_{+}e^{{i \over \hbar}\sqrt{2m(E+V_0)}x}+A_{-}e^{-{i \over \hbar}\sqrt{2m(E+V_0)}x},~0<x<a,\\ 
=C_{+}e^{i{1 \over \hbar}\sqrt{2mE}x},~a<x,
\end{cases}
\end{eqnarray}
where at $x<0$, a right-moving plane wave comes in and a reflected
left-moving wave
of magnitude $B_{-}$  is reflected, and at $x>a$, a right-moving wave of 
magnitude $C_{+}$ is refracted.  Coefficients $A_i$ show the magnitude in the
inside of potential, $0<x<a$. These coefficients are found as \cite{Landau}
\begin{eqnarray}
& &A_{+}=2{1+\sqrt{1+V_0/E} \over (1+\sqrt{1+V_0/E})^2-(1-\sqrt{1+V_0/E})^2 e^{2{i\over \hbar}\sqrt{2m(E+V_0)} a}},\\
& &A_{-}=2{(-1+\sqrt{1+V_0/E})e^{2{i \over \hbar}\sqrt{2m(E+V_0)}a} \over (1+\sqrt{1+V_0/E})^2-(1-\sqrt{1+V_0/E})^2 e^{2{i\over \hbar}\sqrt{2m(E+V_0)} a}},
\end{eqnarray}
and coefficients $B_{-},C_{+}$ are found as 
\begin{eqnarray}
B_{-}&=&1-2{{\sqrt{1+V_0/E}+1+V_0/E+(\sqrt{1+V_0/E}-1-V_0/E)}e^{2{i\over \hbar}\sqrt{2m(E+V_0)} a} \over (1+\sqrt{1+V_0/E})^2-(1-\sqrt{1+V_0/E})^2 
e^{2{i\over \hbar}\sqrt{2m(E+V_0)} a}},\nonumber\\
\\
C_{+}&=&2{(1+\sqrt{1+V_0/E}) +(-1+\sqrt{1+V_0/E})e^{{i \over \hbar}\sqrt{2m(E+V_0)}a} \over (1+\sqrt{1+V_0/E})^2-(1-\sqrt{1+V_0/E})^2 e^{2{i\over \hbar}\sqrt{2m(E+V_0)} a}}\nonumber\\
&\times& e^{{i\over \hbar}(\sqrt{2m(E+V_0)}a-\sqrt{2mE}a)}. 
\end{eqnarray}
Wave packets at $x>0$ and $x<0$ are computed as 
\begin{eqnarray}
\psi_{wp}^{1}&=&\int dk ~N(k) e^{i(kx-\frac{Et}{\hbar})},~x\ll0, \\
\psi_{wp}^{2}&=&\int dk ~N(k) B_{-}(k)e^{-i(kx + \frac{Et}{\hbar})},~x\ll 0, \\
\psi_{wp}^{3}&=&\int dk ~N(k) C_{+}(k) e^{i(kx - \frac{Et}{\hbar})},~x\gg 0,
\end{eqnarray}
where $N(k)$ is given in Eq.~$(\ref{normalization})$.

\begin{figure}[t]
\includegraphics[scale=.4,angle=-90]{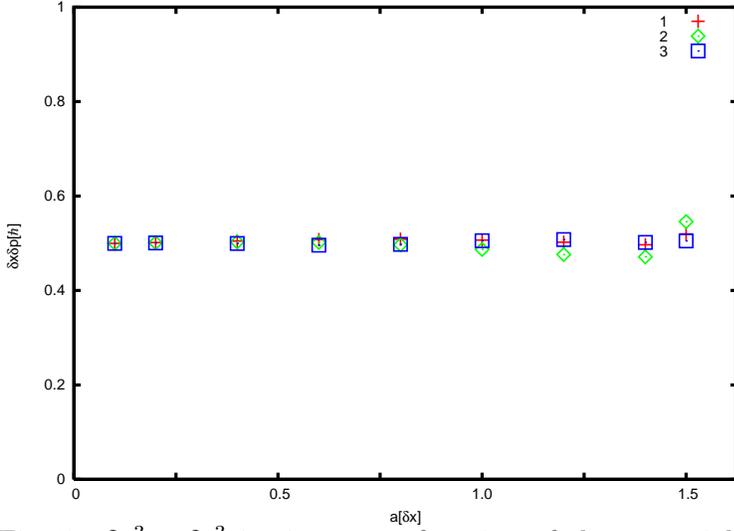}
\refstepcounter{figno}\label{well}
\caption{$\delta
 x^3 \times \delta p^3$ is given as a function of the potential width.
The horizontal axis shows the potential width and
 the vertical axis shows $\delta x \times \delta p$. 1 is for $V_0 = E_0$,~2 is for $V_0 = \frac{3}{2}E_0$,
 and 3 is for $V_0 = \frac{1}{2}E_0$,~where $E_0$ is the average energy
 of the wave packet.}
\end{figure}

Wave packets $\psi_{wp}^{i}(i=1 - 3) $ are computed numerically.
In the negative $x$ region, the minimum wave packet comes in, and
refracted and reflected wave packets are generated. The variances
of  $\psi_{wp}^1$,~$\psi_{wp}^2$, and $\psi_{wp}^3$ are computed
numerically and are given by 
\begin{eqnarray}
& &\delta x^1 \times \delta p^1 = \frac{\hbar}{2},\\
& &\delta x^2  \times \delta p^2 = (1.0 \sim 1.6)\times \frac{\hbar}{2},\\
& &\delta x^3  \times \delta p^3 \simeq \frac{\hbar}{2}.
\end{eqnarray}
The numerical result for $\psi_{wp}^3$ is shown in Fig.~\ref{well},~where
we set $\delta x^1=8.0\times 10^{-10}[\text{m}]$, and the width of the potential
is taken from 0 to $1.5\times \delta x^1$
%one and one-half times spacial size of wavepacket
 and the potential depth is of the order
of the average energy of the wave packet.
In this region,~ $\delta x^3 \times \delta p^3$ is approximately adiabatic
invariant and the minimum wave packet remains nearly minimum after potential
scattering. On the other hand,~$\delta x^2 \times \delta p^2$ becomes
large. 
The large change of $\delta x^2 \times \delta p^2$ from $\delta x^1
\times \delta p^1$ is generated, because there are two boundaries of the
potential barrier/well.
%This is naively because there are two potential shifts for potential barrier/well
%and the reflected wave packet can spread  even in this region by the effect of reflection
%from these boundaries.

%%%%%%%%%%%%%%%%%%%%%%%%%%%% Fig 5 %%%%%%%%%%%%%%%%%%%%%%
%%%%%%%%%%%%%%%%%%%%%%%%%%%% Fig 5 %%%%%%%%%%%%%%%%%%%%%%

\subsection{ Potential scatterings in  three dimensions}
%\subsubsection{One dimensional potentials }

The Schr\"odinger equation with a one-dimensional potential $U( z)$
\begin{eqnarray}
\left[-{{\hbar}^2 \over 2m} {\vec \nabla}^2 +U(z)\right]\psi({\vec x})=E \psi({\vec x})
\end{eqnarray} 
has solutions of the form,
\begin{eqnarray}
\psi({\vec x})=e^{i{\vec p}_T\cdot{\vec x}_T}u(z).
\end{eqnarray}
The first part on the right-hand side is the plane wave in the
transverse direction ${\vec p}_T=(p_x,p_y),{\vec x}_T=(x,y)$ and the 
second  part is the function of $z$.

The wave packet is formed from the above wave function as
\begin{eqnarray}
\psi_{w p}=\int d{\vec k}_T N({\vec k-k_0}_T)e^{i{\vec p}_T\cdot{\vec x}_T}  \int d k_z  N(k_z-k_z^0)u(z),
\end{eqnarray}
where the wave packet in the transverse direction is the wave packet of 
the free wave, and the wave packet in the $z$ direction is the wave
packet of the scattering wave in the one-dimensional potentials studied
in the previous section. 

The wave packet is modified by potentials in the direction of the
 one-dimensional
 potential and is not modified in the transverse directions. 
 The time-dependent wave packet is given by 
  \begin{eqnarray}
\psi({\vec x},t)=\int d{\vec k}_T N({\vec k-k_0}_T) d k_z N(k-k_z^0)
e^{i{\vec p}_T\cdot{\vec x}_T}u(z)e^{-iE({\vec p}_T,p_z)t}.
\end{eqnarray}
Obviously, at small $t$, the momentum width in the transverse direction is 
not modified, 
but the momentum width in the longitudinal direction is modified by the
potential. The change in the width in the longitudinal direction is
similar to that of one dimension.  
%\subsection{short range potentials }

%The Schroedinger equation with a short range potential $U(\vec x)$
%\begin{eqnarray}
%& &[-{{\hbar}^2 \over 2m} {\vec \nabla}^2 +U({\vec x})]\psi({\vec x})
%=E \psi({\vec x})\\
%& &U({\vec x})=0,|{\vec x}| \ge r_c
%\end{eqnarray} 
%has solutions of the form,
%\begin{eqnarray}
%\psi({\vec x})=e^{ipz}+{e^{ipr} \over r}f({\theta}).
%\end{eqnarray}
%The first part in the right hand side is the initial wave and the second 
%part is the scattering wave, which is the outgoing wave from the
%potential center. The time dependent wave function is
% \begin{eqnarray}
%\psi(t,{\vec x})=e^{i(-Et+pz)}+{e^{i(-Et+pr)} \over r}f({\theta})
%\end{eqnarray}  

%\subsection{long range potentials}
%Stationary solution of the Schroedinger equation with long range
%potential has a long tail and coherent properties of wave packets should
%be treated seperately. Coulomb potential and a potential of magnetic
%flux are studied.  
%\subsubsection{Coulomb potential}
%Eigenvalue problems for Coulomb potential are sloved analytically and
%have applications in many area of physics.  So it is not only
%interesting but also
%important to know the coherent properties of Coulomb wave functions.
  
%\subsubsection{AB scattering:magnetic flux potential}
%Two dimensional Schroedinger equation with a vector potential 
%\begin{eqnarray}
%& &[-{1 \over 2m} ({i\hbar \vec \nabla+e{\vec A({\vec x})}})^2 ]\psi({\vec x})=
%E \psi({\vec x})\\
%& &{\vec A}({\vec x})={1 \over r}{\vec e}_{\phi} ,
%\end{eqnarray}

\section{Transformations of wave packets}
In this section, we study the transformations of wave packets and find
changes in wave packets under 
various transformations based on semiclassical treatments. We assume that 
the transition of wave packets is smooth and continuous in
momentum, and naive treatment of the wave packet's parameters is
possible. In these calculations, singular behaviors of the scattering
amplitudes  such as resonances are excluded.  
  
\subsection{Lorentz transformation}
By a Lorentz transformation, a momentum $p_{\nu}$ is transformed to
$p'_{\nu}$ using a matrix  
\begin{eqnarray}
{p_{\mu}}'=\Lambda^{\mu \nu}p_{\nu}. 
\end{eqnarray}
The momentum in the direction of the boost is transformed together with
the energy, but the momentum in the transverse direction is unchanged.
The variances of the momentum components are transformed in the same
manner.

The amplitude for the plane wave is known to 
be covariant under the Lorentz transformation, but the present amplitude 
for the wave packets has a noninvariant part, because the wave packet
size is not fully covariant.  

\subsection{ Addition of potential energy }
In a potential scattering, one particle obtains energy 
from a potential $V_0$
\begin{eqnarray}
E_1=E_2+V_0,
\end{eqnarray}
so wave packet parameters are transformed when a wave packet passes a
potential.  
\subsubsection{ Nonrelativistic case }
For a nonrelativistic particle, momenta are related by 
\begin{eqnarray}
({\vec p}_1)^2+2mV_0=({\vec p}_2)^2.
\end{eqnarray}
By decomposing the momentum vector into the longitudinal component and
the transverse component and their  small deviations,
\begin{eqnarray}
(\delta p_1^l+p_1^{(0)})^2+({\delta {\vec p}_1^T})^2+2mV_0=(\delta p_1^l+p_2^{(0)})^2
+({\delta {\vec p}_1^T})^2,
\end{eqnarray}
we have equalities for the central values and variances of the momenta   
\begin{eqnarray}
& &(p_1^{(0)})^2+2mV_0=(p_2^{(0)})^2,\\
& &2p_1^{(0)}\delta p_1^l=2p_2^{(0)}\delta p_2^l,\\
& &({\delta p_1^T})^2=({\delta p_2^T})^2.
\end{eqnarray}
Thus, the variances of momenta are connected by
\begin{eqnarray}
{\delta p_1^l \over \delta p_2^l}={p_2^{0} \over p_1^{0}}.
\label{mometum-extensions}
\end{eqnarray}
By using the time duration $\tau$, the spatial sizes in the longitudinal  
direction are proportional 
to the central value of the velocity,
\begin{eqnarray}
& &\delta x_1^l =v_1^l \tau, v_1^l={p_1^{0} \over m},\\
& &\delta x_2^l =v_2^l \tau, v_1^l={p_1^{0} \over m},
\end{eqnarray}
hence, the ratio satisfies 
\begin{eqnarray}
{\delta x_1^l \over \delta x_2^l}={p_1^{0} \over p_2^{0}}.
\label{position-extensions}
\end{eqnarray}

From Eqs.~({\ref{mometum-extensions}}) and
 ({\ref{position-extensions}}),
 the product of momentum extensions and position extensions 
becomes adiabatic invariant, 
\begin{eqnarray}
{\delta x_1^l \delta p_1^l \over \delta x_2^l \delta p_2^l}=\text{constant}.
\label{position-momentum-extensions}
\end{eqnarray}
The  momentum extensions and spatial sizes in the transverse direction
are unchanged 
\begin{eqnarray}
\delta x_1^T ={\hbar \over \delta p_1^T}={\hbar \over \delta p_2^T}{\delta p_2^T \over \delta p_1^T}=\delta x_2^T
\label{ratio-widths-nonrelativistic-T}
\end{eqnarray}
and are proportional to their energies. 
\subsubsection{ Relativistic case }
For the relativistic particle, the relation is modified to 
\begin{eqnarray}
\sqrt{(\delta p_1^l+p_1^{(0)})^2+({\delta {\vec p}_1^T})^2+
m^2}=\sqrt{(\delta p_2^l+p_2^{(0)})^2+({\delta {\vec p}_2^T})^2+m^2}+V_0.
\end{eqnarray}
Thus, we have 
\begin{eqnarray}
& &\sqrt{(p_1^{(0)})^2+m^2}=\sqrt{(p_2^{(0)})^2+m^2}+V_0,\\
& &2{p_1^{(0)} \over E_1}\delta p_1^l=2{p_2^{(0)} \over E_2}\delta p_2^l,\\
& & ({\vec \delta p_1}^T)^2=({\vec \delta p_2}^T)^2,
\end{eqnarray}
hence, the spatial sizes of extensions in the longitudinal direction 
are given by
\begin{eqnarray}
\delta x_1^l ={\hbar \over \delta p_1^l}={\hbar \over \delta p_2^l}
{\delta p_2^l \over \delta p_1^l}={{p_1^l \over E_1} \over {p_2^l \over E_2}}
 \delta x_2^l.
\end{eqnarray}
The  momentum extensions and spatial sizes in the transverse direction
are unchanged 
\begin{eqnarray}
\delta x_1^T ={\hbar \over \delta p_1^T}={\hbar \over \delta p_2^T}{\delta p_2^T \over \delta p_1^T}=\delta x_2^T.
\label{ratio-widths-relativistic-T}
\end{eqnarray}

In the low-energy region, the energy $E_1$ and $E_2$ are $mc^2$  and 
the relation of the spatial extensions coincides with that of
the nonrelativistic case, Eq.~$(\ref{ratio-widths-nonrelativistic-T})$.  
If both momenta are relativistic, the velocities are almost $c$,
\begin{eqnarray}
{p_1 \over E_1}={p_2 \over E_2}=c
\end{eqnarray}
and both sizes are almost the same,
\begin{eqnarray}
\delta x_1 = \delta x_2.
\end{eqnarray}
The massless particle has a light velocity and does not spread  in the
direction of motion, and the massive
particle has the same property, that is, the coherence length is not 
transformed in the relativistic regime.  On the other hand, the massive
particle expands when its energy is enlarged by the potential energy from 
the nonrelativistic region to the relativistic region and has a size at
light velocity, 
\begin{eqnarray}
\delta x_c={c \over v}\delta x_v
\end{eqnarray}
if it has $\delta x_v$ at velocity $v$.

\subsection{ Scale transformation}
In scale transformation, a momentum is multiplied by a constant factor $\lambda$,
\begin{eqnarray}
{ p_{\mu}}_2=\lambda  {p_{\mu}}_1,
\end{eqnarray}
which is consistent with the energy and momentum relation of the
massless particle. Thus, this transformation is applied only to the massless
particle.

The variance is transformed then by
\begin{eqnarray}
\delta {p_{\mu}}_2=\lambda \delta {p_{\mu}}_1,
\end{eqnarray}
hence, the spatial sizes of extensions are given by
\begin{eqnarray}
\delta {x_{\mu}}_2 ={1 \over \lambda } \delta {x_{\mu}}_1.
\end{eqnarray}
For  $\lambda << 1$, we have
\begin{eqnarray}
\delta x_2 >> \delta x_1.
\end{eqnarray}

\section{Refraction and reflection }

We study situations where a half space is occupied by one medium and
another half is occupied by another medium. A wave packet in one half is
reflected at the other half and refracted at the boundary.  Wave packets
of these situations are studied here. 

In the situation where one half is  the vacuum and another half is
filled with medium, the wave in  the medium has a mean 
free path. Thus, the wave  that is produced in the medium first and
emitted into the vacuum later is described using a wave of finite coherence 
length. Although the   wave in the vacuum is  described using the free 
Hamiltonian, this wave is  the wave
packet of having a finite mean free time and a finite 
energy width.   
Thus, 
the sizes of wave packets in vacuum are determined using the mean free
path in the medium if the wave is produced in the medium and emitted
into the vacuum.   
\subsection{Electrons from metal to vacuum}
Electrons in the metal follow  energy dispersions that are characteristic
of  the band structure and have a lower energy than that in the vacuum
because of the value of the work function. The energy dispersion is approximately
expressed  by a
quadratic form of an effective mass $m_\text{eff}$ that is different from 
the mass in the vacuum  $m_0$.   For a spherically 
symmetric band, we have  the relation of energies  
between the momentum in metal ${\vec p}_1$ and the momentum in vacuum
${\vec p}_2$,
\begin{eqnarray}
E_0+{{\vec p_1}^2 \over 2m_\text{eff}}=  {{\vec p_2}^2 \over 2m_{0}},
\end{eqnarray}
where $E_0$ is the work function in the metal.

By decomposing the momenta into the components and the central values
 $p_i^{0}$ and deviations in the longitudinal direction, $\delta p_l$,
 and 
those in the transverse directions, ${\vec \delta p}_i$, we have
\begin{eqnarray}
& &E_0+{{p_1^{0}}^2 \over 2m_\text{eff}}=  {{ p_2^{0}}^2 \over 2m_{0}},\\
& &{2 {p_1^{0}}\delta p_1^{l} \over 2m_\text{eff}}=  {{ p_2^{0}}\delta p_2^{l} \over 2m_{0}},\\
& &{{\delta {\vec p}_1}^2 \over 2m_\text{eff}}=  {{\delta{\vec  p}_2}^2 \over 2m_{0}}.
\end{eqnarray}
Electrons propagate in the form of wave packets of the above parameters.
\subsection{Lights from medium to vacuum}

\subsubsection{ Without absorption}

In an insulator medium, the dielectric constant is different from that
in a vacuum and is real if there is no absorption. In this
situation, a momentum in the medium ${\vec p}_1$ and a momentum in the
vacuum ${\vec p}_2$ are connected by   
\begin{eqnarray}
& &c_m { p}_1=c { p}_2,\\
& &c_m=c\sqrt{\mu_0 \epsilon_0 \over \mu \epsilon},
\end{eqnarray}
where $\mu$ and $\mu_0$ are the  permeabilities of the medium and
vacuum and $\epsilon$ and $\epsilon_0$ are the dielectric constants of the
medium and vacuum, respectively. $c_m$ is the light velocity in the medium and $c$
is the light velocity in the vacuum.   
\subsubsection{ Finite absorption}
In a system of absorption, the photon energy has an imaginary
part
\begin{eqnarray}
\Gamma=\mu {1 \over \rho},
\end{eqnarray}
where $\rho$ is resistivity.
The relation of the momenta at the boundary is given by
\begin{eqnarray}
& &c_m { p}_1=c{ p}_2,\\
& &c_m=c\sqrt{\mu_0 \epsilon_0 \over \mu \epsilon},
\end{eqnarray}
but owing to the imaginary part of the energy, the wave lives for a finite 
time $\tau$
 \begin{eqnarray}
\tau={\epsilon  \rho }
\end{eqnarray}   
in the medium. Thus, the light that is emitted from the medium into the
vacuum has an uncertainty  of energy $\Delta E$,
\begin{eqnarray}
\Delta E={\hbar \over \tau}.
\end{eqnarray}
Lights propagate in the form of wave packets of the above parameters.
%\section{Mean free path }

%\section{Coherence and  scattering amplitude}

%\section{Symmetries  }

%Symmetric coherence 

%\section{Applications}

%\subsection{neutrino oscillation} 

%\subsection{extremely high energy cosmic ray scattering with background 
%radiation}

\section{Many-body processes}
The coherence length of a particle in final states of  many-body processes 
is  determined on the basis of the uncertainties
of the   energy and momentum of the initial states. The momentum correlation
(Eq.~(\ref{correlation-wave packet})) of a one-particle state is used for
obtaining the coherence length. 
\subsection{Two-body decay}
In a two-body decay, $A \rightarrow B+C$, the energy and magnitude of 
momentum of B
and C are fixed in the rest system of A. The correlation function
(Eq.~$(\ref{correlation-wave packet})$) is defined as
\begin{eqnarray}
C({\vec p}_1,{\vec p}_2)=\sum_{{\vec p}_C} \langle {\vec p}_1,{\vec p}_C |T|A \rangle (\langle {\vec p}_2,{\vec p}_C |T|A \rangle)^* ,
\end{eqnarray}
where the above amplitude is proportional to the amplitude $\tilde T$
and the delta function of energy momentum conservation,
\begin{eqnarray}
\langle {\vec p}_1,{\vec p}_C |T| A \rangle =\delta^{4}({p}_A-{p}_1-{p}_C) \tilde T
\end{eqnarray}
when the state $A$ is the eigenstate of the energy and momentum. Thus, 
if the state $A$ is the eigenstate of the energy and momentum, the
above correlation function becomes proportional to 
\begin{eqnarray}
  C({\vec p}_1,{\vec p}_2)=\tilde T {\tilde T}^{*}\delta^{4}({p}_1-{ p}_2).
\end{eqnarray}
On the other hand, when the state $A$ is a wave packet of the function 
$F({\vec p}_A)$, the correlation function is given by  
\begin{eqnarray}
C({\vec p}_1,{\vec p}_2)&=&\sum_{{\vec p}_C} \int d {\vec p}_A d {\vec p}_{A'}
\langle {\vec p}_1,{\vec p}_C |T|{\vec p}_A \rangle 
(\langle {\vec p}_2,{\vec p}_C |T|{\vec p}_A' \rangle)^* F({\vec p}_A) F^{*}({\vec p}_{A'})\nonumber \\
&=&\int d{\vec p}_A d{\vec p}_{A'} \delta({\vec p}_1-{\vec p}_2-{\vec p}_A+
{\vec p}_{A'})F({\vec p}_A) F^{*}({\vec p}_{A'})\tilde T {\tilde T}^{*} \nonumber\\
& &\times \delta({p}_A^{0}-{p}_1^{0}-{p}_C^{0})\delta({p}_A^{0}-{p}_1^{0}-
{p}_C^{0})\nonumber \\
&=&\int d{\vec p}_A F({\vec p}_A) F^{*}({\vec p}_{A}-{\vec p}_2+{\vec p}_1)
\tilde T {\tilde T}^{*} \nonumber\\
& &\times \delta({p}_A^{0}-{p}_1^{0}-{p}_C^{0})\delta({p}_A^{0}-{p}_1^{0}-
{p}_C^{0}).
\end{eqnarray}
Thus, the momentum correlation $C({\vec p}_1,{\vec p}_2)$ is determined using
the momentum distribution function $F({\vec p}_A)$ of the initial state.
\subsection{Three-body decay}
In a three-body decay, $A \rightarrow B+C+D$, the energy and magnitude of 
momentum of B, C and D vary even in the rest system of A. If
particle B is measured and the other states are not measured but 
summed, the result of the correlation function for $B$ is the same
as that of two-body decay. The correlation function is given by
\begin{eqnarray}
C({\vec p}_1,{\vec p}_2)=\sum_{{\vec p}_C,{\vec p}_D} \langle {\vec p}_1,{\vec p}_C,{\vec p}_D |T|{\vec p}_A \rangle 
(\langle {\vec p}_2,{\vec p}_C,{\vec p}_D |T|{\vec p}_A \rangle)^* ,
\end{eqnarray}
and  is proportional to the delta function if the state $A$ is the
eigenstate of the energy and momentum,  
\begin{eqnarray}
  C({\vec p}_1,{\vec p}_2)=\delta^{4}({ p}_1-{ p}_2)\tilde T {\tilde T}^{*}.
\end{eqnarray}
On the other hand, when the state $A$ is a wave packet of the function 
$F({\vec p}_A)$, the correlation function is given by  
\begin{eqnarray}
C({\vec p}_1,{\vec p}_2)&=&\sum_{{\vec p}_C,{\vec p}_D} \int d {\vec p}_A d {\vec p}_{A'}
\langle {\vec p}_1,{\vec p}_C,{\vec p}_D |T|{\vec p}_A \rangle 
(\langle {\vec p}_2,{\vec p}_C,{\vec p}_D |T|{\vec p}_{A'} \rangle)^* F({\vec p}_A) F^{*}({\vec p}_{A'})\nonumber \\
&=&\int d{\vec p}_A d{\vec p}_{A'} \delta({\vec p}_1-{\vec p}_2-{\vec p}_A+
{\vec p}_{A'})F({\vec p}_A) F^{*}({\vec p}_{A'})\tilde T {\tilde T}^{*} \nonumber\\
&=&\int d{\vec p}_A F({\vec p}_A) F^{*}({\vec p}_{A}+{\vec p}_2-{\vec p}_1)\tilde T {\tilde T}^{*}.
\end{eqnarray}
Thus, the momentum correlation $C({\vec p}_1,{\vec p}_2)$ is determined by
the momentum distribution function $F({\vec p}_A)$ of the initial state $A$.

\subsection{Two-body collision }
The coherence lengths of collision products are treated in the same manner as
the decay products of the previous section and are determined on the
basis of the 
uncertainties of the energy and momentum of the initial states. We study 
the momentum correlations (Eq.~$(\ref{correlation-wave packet})$) of one 
particle also in the collision products.
In a two-body collision, $A +B \rightarrow C+D$, we study 
the correlation function
(Eq.~$(\ref{correlation-wave packet})$) defined as
\begin{eqnarray}
C({\vec p}_1,{\vec p}_2)=\sum_{{\vec p}_C} \langle {\vec p}_1,{\vec p}_C |T|A,B \rangle (\langle {\vec p}_2,{\vec p}_C |T|A,B \rangle)^* ,
\end{eqnarray}
where the above amplitude is proportional to the amplitude $\tilde T$
and the delta function of energy momentum conservation,
\begin{eqnarray}
& &\langle {\vec p}_1,{\vec p}_C |T| A,B \rangle =\delta^{4}({p}_\text{initial}-{p}_1-{p}_C) \tilde T,\\
& &p_{\text{initial}=p_A+p_B}\notag
\end{eqnarray}
when states $A$ and $B$ are the eigenstates of the energy and momentum. Thus, 
if the states $A$ and $B$ are the eigenstates of the energy and momentum, the
above correlation function becomes proportional to 
\begin{eqnarray}
  C({\vec p}_1,{\vec p}_2)=\tilde T {\tilde T}^{*}\delta^{4}({p}_1-{ p}_2).
\end{eqnarray}

On the other hand, when states $A$ and $B$ are  wave packets
of finite spreads, the wave functions overlap within  a finite space-time
region,
 \begin{eqnarray}
\Delta t_\text{overlap} \neq \infty ,~\Delta {\vec x} \neq \infty
\end{eqnarray}
and the energy-momentum conservation is slightly violated,
\begin{eqnarray}
 \Delta E =E_i-E_f\neq 0,~ \Delta {\vec p} ={\vec p}_i-{\vec p}_f\neq 0.
\end{eqnarray}
The correlation function is expressed using the wave  
functions $F({\vec p}_A)$ and $F({\vec p}_B)$ as
\begin{eqnarray}
C({\vec p}_1,{\vec p}_2)&=&\sum_{{\vec p}_C} \int d {\vec p}_A d {\vec p}_{A'}
 \int d {\vec p}_B d {\vec p}_{B'}
\langle {\vec p}_1,{\vec p}_C |T|{\vec p}_A,{\vec p}_B \rangle 
(\langle {\vec p}_2,{\vec p}_C |T|{\vec p}_{A'},{\vec p}_{B'} \rangle)^* 
\nonumber \nonumber\\
&\times &F({\vec p}_A) F^{*}({\vec p}_{A'})F({\vec p}_B) F^{*}({\vec p}_{B'})\nonumber\\
&=&\int d{\vec p}_A d{\vec p}_{A'} d{\vec p}_B d{\vec p}_{B'} 
\delta({\vec p}_1-{\vec p}_2-{\vec p}_A+
{\vec p}_{A'}-{\vec p}_B+{\vec p}_{B'} )\nonumber\\
&\times &F({\vec p}_A) F^{*}({\vec p}_{A'})
F({\vec p}_B) F^{*}({\vec p}_{B'})\tilde T {\tilde T}^{*} \nonumber\\
&=&\int d{\vec p}_A F({\vec p}_A)d{\vec p}_B F({\vec p}_B) F^{*}({\vec p}_{A}-{\vec p}_2+{\vec p}_1)F^{*}({\vec p}_{A}-{\vec p}_2+{\vec p}_1)
\tilde T {\tilde T}^{*}.
\end{eqnarray}
Thus, the momentum correlation $C({\vec p}_1,{\vec p}_2)$ is determined using
the wave functions of the initial
state. This result is also applied to  many-body scatterings. The
correlation function $C({\vec p}_1,{\vec p}_2)$ was used in
\S2.2.

\section{Summary}
In this paper, we  
showed  that one-particle states  are described using wave packets of finite
coherence lengths, i.e., finite wave packet sizes in various
situations. The wave packet size is determined
either from a one-particle effect or from a many-particle effect. In the former,
a finite mean free path is the origin of the wave packet. The finite mean
free path makes one particle have a finite spatial extension and
a finite momentum uncertainty. The state of a finite mean free path 
is a nonstationary  state and is varied with  time.  The
state is extended also in energy, and the energy width is determined
either from the mean free path or from 
the mean free time. Two values  are consistent with each other.   In the
latter, a one-particle state is generated as  a superposition of plane waves owing to
many particle effects and has a correlation with a wave packet. The
situation is similar to the fact that a
spherical wave 
is produced by a short range potential. The spherical wave is a
superposition of plane waves of different orientations. Usually, a
particle is detected
using a detector of finite size, and the number of events is determined
separately at different angles, so it is difficult to observe directly the coherence
of different angles. To test the coherence of different
angles directly, a particular detector that responds in a wide
orientation may be necessary. 

We verified in the latter sections that once particles of finite coherence are
produced, the finite coherence propagates and transmits  to other
particles due to
scatterings and many-body effects.

  In the next paper, we study various applications of wave packets in 
interference phenomena in large-scale physics.\\
\ \\
%\section{References}
    
  %%%%%%%%%%%%%%%%%%%%%%%%%%%%%%%%%%%%%%%%%%%%%%%%%%%%%%%%%%%%%%%%%%%%%%%%%%%
%\newpage
\section*{Acknowledgements}

This work was partially supported by a Grant-in-Aid
for Scientific Research (Grant No. 19540253)
and a Grant-in-Aid for Scientific Research on 
Priority Area (Progress in Elementary Particle Physics of the 21st
Century through Discoveries  of Higgs Boson and Supersymmetry, Grant 
No. 16081201) provided by 
the Ministry of Education, Culture, Sports, Science, and Techonology, Japan. 
\\
%%%%%%%%%%%%%%%%% Reference %%%%%%%%%%%%%%%%%%%%%%%%%%%%%%%%
%Bibtex'»È'¤Ã'¤¿'¸å'¡¢'½¤'Àµ'¤·'¤Æ'¥Õ'¥¡'¥¤'¥ë'¤ò'Æɒ¤ß'¹þ'¤à
%
%\bibliographystyle{unsrt}
%\bibliography{ref}
%\input{coherence-length-3.bbl}

%{\bf Appendix A:Mean free path in matter}

%Mean free paths of the electron, photon, muon, proton, and
%pi-meson, and neutron in various situations are determined by their
%interactions with matters. It is important to know the orders of their 
%values. So 
%table of mean free paths in the
%medium are given in the followings:
%\begin{eqnarray}
%& &l_{e}=  \\
%& &l_{p}=  \\
%& &l_{\mu}= \\
%& &l_{\pi}=\\
%& &l_{n}=.
%\end{eqnarray}

%The values vary a lot depending on  particles and medium.

\appendix
\section{Cross Sections around  the Decoupling Time}

Around the decoupling time of the early universe, the densities of photon, 
electron, and proton are given by 
\begin{eqnarray}
n_p&=&n_e=4 \times 10^{17}~[\text{m}^{-3}],  \\
n_{\gamma}&=& 10^9 \times n_p.
\end{eqnarray}
Scattering cross sections of Thomson scattering and Rutherford
scattering are given by
\begin{eqnarray}
\sigma_\text{Ru}&=&4\pi ({e^2 \over 4\pi\epsilon_0 mv^2})^2 \log
 \Lambda=4.4\times 10^{-17}[\text{m}^2] \times \log \Lambda, \\
& &\Lambda={\gamma \hbar v 4\pi \epsilon_0\over e^2},~ mv^2=kT, \nonumber \\
\sigma_\text{Th}&=&{ 8 \pi r_e^2 \over 3}=0.6 \times 10^{-28}~[\text{m}^2] ,\\
& &r_e={e^2 \over 4\pi \epsilon_0 m_e c^2}, \nonumber
\end{eqnarray}
where the cutoff parameter $\log \Lambda=10$ and the temperature
$T=3000$~K are used.
Hence, the mean free paths are  given by
\begin{eqnarray}
& &l_\text{Ru}=5.7\times 10^{-3}~[\text{m}],\\
& &l_\text{Th}=5~[\text{m}],
\end{eqnarray}
and we have 
\begin{eqnarray}
N_\text{T}={l_\text{Th} \over l_\text{Ru}}=10^3.
\end{eqnarray}
\section{Total Momentum Uncertainty of N Particles}

When a particle is surrounded by \textit{N} particles and  interacts with them
coherently, one particle in the final state obtains a total momentum
uncertainty from \textit{N} particles. The total amplitude of this processes is
written as
\begin{eqnarray}
f=\int \prod_i d{\vec p}_i F({\vec p}_i)
\langle {\vec q}_1,\cdots, {\vec q}_M,|T|{\vec p}_1,\cdots {\vec p}_N \rangle.
\end{eqnarray}   
We study the case where all the particle states of  the initial state are
described using the
same wave packet for simplicity. In other cases, the following
conclusion is  the same. 

The product of \textit{N} Gaussian functions 
\begin{eqnarray}
& &F({\vec p}_1)F({\vec p}_2)\cdots F({\vec p}_N),\\
& & F({\vec p}_1)=Ne^{-\sigma ({{\vec p}_1-{\vec p}_1^{0}})^2}
\end{eqnarray}
is decomposed into the function of the total momentum ${\vec p}_T$ and 
relative momenta ${\vec p}_r$ as 
\begin{eqnarray}
F_T({\vec p}_T-{\vec p}_T^{0})  \prod_r F_r({\vec p}_r-{\vec p}_r^{0}).
\end{eqnarray}
In the above function, $F_T({\vec p}_T-{\vec p}_T^{0} )$ is given by
\begin{eqnarray}
& &F_T({\vec p}_T-{\vec p}_T^{0})= \tilde N e^{-{\sigma \over N}({\vec p}_T-{\vec p}_T^{0})^2},  \\
& &{\vec p}_T=\sum_i{\vec p}_i,~{\vec p}_T^{0}=\sum_i{\vec p}_i^{0},
\end{eqnarray}
where $\tilde N$ is a normalization constant.  Thus, the spread of the 
total momentum increases with $N$ and becomes ${\sqrt N}$ times the 
spread of one particle. These \textit{N} particles give the momentum
uncertainty ${\sqrt N} \Delta p$ to the particle.  

\section{ A Wave Function of Statistical Model }

The many-body wave function that has a minimum uncertainty of field operators
$\phi({\vec x})$ and $\pi({\vec x})$  for a Boson
\begin{eqnarray}
& &\delta \phi_g^2 \delta \pi_g^2 \geq {\hbar^2 \over 4}\int  d{\vec p}g^2({\vec p}),
\\
& &\delta \phi_g^2=\langle  \phi_g^2\rangle-\langle \phi_g\rangle^2,~\delta \pi_g^2=\langle  \pi_g^2\rangle-\langle \pi_g\rangle^2,~\nonumber \\
& &\phi_g=\int d{\vec p}\phi({\vec p})g({\vec p}),\\
& &\pi_g= \int d{\vec p}\pi({\vec p})g({\vec p}),         
\end{eqnarray}  
is a coherent state  
\begin{eqnarray}
& &|\psi \rangle=Ne^{\int d{\vec p} g({\vec p})a^{\dagger}({\vec p}) }|0\rangle,\\
& &N^2=e^{-\int d{\vec p} |g({\vec p})|^2}.
\end{eqnarray}
This coherent state satisfies 
\begin{eqnarray}
& &\langle \psi| a({\vec q})|\psi \rangle=g({\vec q}), \\
& &\langle \psi |a({\vec q})^{\dagger} a({\vec q})|\psi \rangle
=| g({\vec q})|^2,
\end{eqnarray}
and the number density agrees with that of our
statistical model if 
\begin{eqnarray}
|g({\vec p})|^2=n_P({\vec p}).
\end{eqnarray} 
An example of the weight is 
\begin{eqnarray}
g({\vec p},{\vec X})=\sqrt {n_P({\vec p})}e^{-i{\vec p}\cdot{\vec X}},
\end{eqnarray} 
where we choose a suitable vector ${\vec X}$.
For Fermion $b^{\dagger}({\vec p})$, a wave function is chosen in the same
manner, and we have  
\begin{eqnarray}
|\Psi \rangle=N \prod_i e^{\int d{\vec p} (f({\vec p},{\vec X}_i) a^{\dagger}({\vec p}) + g({\vec p},{\vec X}_i)b^{\dagger}({\vec p})  )}|0 \rangle. 
\end{eqnarray}

{}

\end{document}